\documentclass[unsortedaddress,nofootinbib,11pt]{revtex4}
\usepackage[utf8]{inputenc}
\usepackage{amsmath,empheq,color}
\usepackage{amsfonts}
\usepackage{amsthm}
\usepackage{amssymb}
\usepackage{graphicx}
\usepackage{hyperref}
\usepackage[normalem]{ulem}
\usepackage{epigraph}
\usepackage{mathrsfs}
\usepackage{bbm}
\usepackage{lineno}
\usepackage{scalerel,stackengine}

\usepackage{caption}
\usepackage{subcaption}
\begin{document}

\title{Squeeze operator: a classical view}

\author{Angel Garcia-Chung}
\email{alechung@xanum.uam.mx} 


\affiliation{Departamento de F\'isica, Universidad Aut\'onoma Metropolitana - Iztapalapa\\
San Rafael Atlixco 186, Ciudad de M\'exico 09340, M\'exico}

\begin{abstract}
In this paper we show the relation between $sp(4,\mathbb{R})$, the Lie algebra of the symplectic group, and the elements of the symplectic group $Sp(4,\mathbb{R})$. We use this relation to provide a classical analog of the squeeze operator $\widehat{S}(\zeta)$. This classical squeeze matrix shares some similarities with the correlation matrix ${\bf V}^{(2)}$ and its amount of squeezing is half of that in the correlation matrix.
\end{abstract}

\maketitle

\section{Introduction}

The squeezed states are broadly used in many areas of physics \cite{walls1983squeezed, schnabel2017squeezed, walls2007quantum, adesso2014continuous, braunstein2005quantum}. An arena where these states are important is in quantum cosmology \cite{gasperini1993quantum, grain2019squeezing}, particularly in Loop Quantum Cosmology (LQC) \cite{ashtekar2003mathematical, bojowald2010canonical, bojowald2011quantum}. In this scenario, squeezed states for a single mode show some of the relevant features of the quantum bounce \cite{mielczarek2012gaussian, gazeau2013quantum, diener2014numerical, ashtekar2006quantum, ashtekar2006quantum2}.  Recently, the entanglement between loop quantum cosmology and matter has been considered \cite{husain2019matter} using discrete variables in the matter sector. However, the analysis for the two-mode squeezed states with continuous variables in LQC, where entanglement between the degrees of freedom of space and matter can occur, is still absent.

The squeezed states result from the action of the squeeze operators on coherent states or in the vacuum state \cite{walls2007quantum, adesso2014continuous, braunstein2005quantum}. Usually, the squeeze operators are considered in the Fock representation, which is based on annihilation and creation operators, or in the Wigner representation, using the Wigner functional. However, in models like in LQC, where the representation of the observables is not weakly continuous, these schemes are not necessarily suitable.  Instead, the Schr\"odinger representation, which is the scheme inherited from the quantization procedure, seems the  natural scheme to be considered \cite{ashtekar2003quantum, corichi2007polymer, flores2013propagators, velhinho2007quantum, pawlowski2014separable, Garcia-Chung:2020cag}. For this reason, the analysis of the squeeze operator for bi-partite systems in the Schr\"odinger representation is needed. 

The Schr\"odinger representation of the squeeze operator for a bi-partite system can be obtained using the representation of the squeeze generator operator via the exponential map \cite{hall2018theory}. In the case of LQC this cannot be done, let us clarify why. Consider the squeeze operator for a bi-partite system $\widehat{S}(\zeta)$, which can be written as $\widehat{S}(\zeta) = e^{ \widehat{s}(\zeta)}$. Here, $\widehat{s}(\zeta)$ is the squeeze generator and $\zeta$ is a complex number  labelling the amount of squeezing. In order to obtain the Schr\"odinger representation for $\widehat{S}(\zeta)$ we may first calculate the Schr\"odinger representation of $\widehat{s}(\zeta)$, and then, via its exponential map, we can obtain the representation of $\widehat{S}(\zeta)$. Apart from the mathematical challenge that this operation requires, this is not possible in LQC due to there is no momentum operator \cite{ashtekar2003quantum, corichi2007polymer}, that is to say, there is no infinitesimal generator $\widehat{s}(\zeta)$. Therefore, another approach is needed, essentially an approach in which the Schr\"odinger representation of $\widehat{s}(\zeta)$ is not a key ingredient of the construction. This approach is the unitary representation of the symplectic group. 

The unitary representation of $Sp(2n, \mathbb{R})${\footnote{The metaplectic group $Mp(2n, \mathbb{R})$ is a double cover of the symplectic group $Sp(2n, \mathbb{R})$. Hence, its representation is usually related with the representation of the symplectic group.}}, was given in \cite{moshinsky1971linear} and revisited in \cite{wolf2016development, torre2005linear}. In \cite{moshinsky1971linear} the representation is carried out using the entire group via their integral representation. This gave rise to what is nowadays called ``quadratic Fourier transforms'', see \cite{wolf2016development, torre2005linear} for more details. As a consequence, this representation allows us to represent all the elements in $Sp(4, \mathbb{R})$ and not just those that are connected to the identity element of the group \cite{hall2018theory}. In order to consider the unitary representation of the squeeze operator we are force to consider the symplectic matrix whose unitary representation gives rise to the squeeze operator $\widehat{S}(\zeta)$. To do so, we need (i) to find the relation between the Lie generator $\widehat{s}(\zeta)$ and its classical analog ${\bf m}_{\bf s}$ and (ii) to derive the classical symplectic matrix ${\bf M}_{\bf s}$ such that it is the exponential map of ${\bf m}_{\bf s}$, i.e., ${\bf M}_{\bf s} = e^{ {\bf m}_{\bf s}}$. To the best of our knowledge, there is no reference in the literature where the calculation of the exponential map ${\bf M} = e^{ {\bf m}}$ for arbitrary matrices ${\bf m}$ has been carried out in detail for $Sp(4, \mathbb{R})$. 

For  these reasons, the main purpose of the present work is to provide the detailed calculation of the exponential map ${\bf M} = e^{ {\bf m}}$ for an arbitrary element ${\bf m} \in sp(4,\mathbb{R})$ of the Lie algebra of $Sp(4, \mathbb{R})$.  Additionally, we  apply the result to ${\bf m}_{\bf s}$ and ${\bf M}_{\bf s}$, where ${\bf M}_{\bf s}$ is what we call the classical squeeze operator.

This paper is organized as follows: in section (\ref{LAAnalysis}) we show the details of the calculation of ${\bf M} = e^{ {\bf m}}$ and in section (\ref{ClassicalAnalysis}) we discuss the classical squeeze operator ${\bf M}_{\bf s}$. In section (\ref{Conclusion}) we give the conclusions.

\section{Lie algebra and group analysis} \label{LAAnalysis}

The symplectic group $Sp(4,\mathbb{R})$ is given by $4\times 4$ matrices ${\bf M}$ for which the following condition holds
\begin{equation}
\left( \begin{array}{cc} {\bf J} & 0 \\ 0 & {\bf J} \end{array} \right) = {\bf M} \left( \begin{array}{cc} {\bf J} & 0 \\ 0 & {\bf J} \end{array} \right) {\bf M}^T ,\label{SpCond}
\end{equation}
\noindent where ${\bf M}^T$ is the transpose matrix and ${\bf J}$ is a $2\times2$ matrix given as
\begin{equation}
{\bf J} = \left( \begin{array}{cc} 0 & 1 \\ -1 & 0 \end{array} \right).
\end{equation}

Matrix ${\bf M}$ can be written as
\begin{equation}
{\bf M} :=  \left( \begin{array}{cc} {\bf A} & {\bf B} \\ {\bf C} & {\bf D} \end{array}\right),
\end{equation}
\noindent where the block matrices ${\bf A}$, ${\bf B}$, ${\bf C}$ and ${\bf D}$ satisfy the conditions 
\begin{equation}
{\bf J} = {\bf A} {\bf J} {\bf A}^T +  {\bf B} {\bf J} {\bf B}^T =  {\bf C} {\bf J} {\bf C}^T +  {\bf D} {\bf J} {\bf D}^T , \qquad {\bf 0} =  {\bf A} {\bf J} {\bf C}^T +  {\bf B} {\bf J} {\bf D}^T ,
\end{equation}
\noindent which are a result of (\ref{SpCond}). 

The Lie algebra of $Sp(4,\mathbb{R})$, denoted as $sp(4,\mathbb{R})$, is given by $4\times4$ matrices ${\bf m}$ of the form
\begin{equation}
{\bf m} = \left( \begin{array}{cc} {\bf J} & 0 \\ 0 & {\bf J} \end{array} \right) {\bf L},
\end{equation}
\noindent where the matrix ${\bf L}$ is a real symmetric matrix written as
\begin{equation}
{\bf L}= \left( \begin{array}{cc} {\bf a} & {\bf b} \\ {\bf b}^T & {\bf c} \end{array}\right),
\end{equation}
\noindent and ${\bf a}$ and ${\bf c}$ are $2\times2$ symmetric matrices and ${\bf b}$ is a $2\times2$ matrix. 

Due to $Sp(4, \mathbb{R})$ is a Lie group, some of its elements can be obtained via the exponential map \cite{hall2018theory} of the Lie algebra element ${\bf L}$ as
\begin{equation}
{\bf M} = \exp{ \left[ \left( \begin{array}{cc} {\bf J} & 0 \\ 0 & {\bf J} \end{array} \right) {\bf L} \right]}.  \label{MandS}
\end{equation}
\noindent The aim of this section is to obtain the relation between the block matrices ${\bf A}$, ${\bf B}$, ${\bf C}$ and ${\bf D}$ and the Lie algebra element ${\bf L}$. What we will obtain is a relation between the block matrices ${\bf a}$, ${\bf b}$ and ${\bf c}$ of the Lie algebra element ${\bf L}$ and the matrices  ${\bf A}$, ${\bf B}$, ${\bf C}$ and ${\bf D}$ of the group element ${\bf M}$.

To proceed, let us expand the exponential in (\ref{MandS}) and collect together the even and odd terms of the expansion as follows
\begin{eqnarray}
{\bf M} &=& \left[  {\bf 1} + \frac{1}{2!} {\bf S} + \dots + \frac{1}{(2n)!} {\bf S}^n + \dots 	\right] +  \sqrt{{\bf S}} \left[ {\bf 1} + \frac{1}{3!}  {\bf S} + \dots + \frac{1}{(2n+1)!} {\bf S}^n + \dots   \right], \label{Expansion} 
\end{eqnarray}
\noindent where the matrix ${\bf S}$ is defined as
\begin{equation}
{\bf S} = \left[ \left( \begin{array}{cc} {\bf J} & 0 \\ 0 & {\bf J} \end{array} \right) \left( \begin{array}{cc} {\bf a} & {\bf b} \\ {\bf b}^T & {\bf c} \end{array} \right) \right]^{2} = \left( \begin{array}{cc} - (\det {\bf a} + \det {\bf b}) {\bf 1}_{2\times2} & {\bf J} {\bf d} \\ - {\bf J} {\bf d}^T & - (\det {\bf b} + \det {\bf c}) {\bf 1}_{2\times2} \end{array}\right), \label{SDefinition}
\end{equation}
\noindent and the matrix ${\bf d}$ is given by ${\bf d} = {\bf a} {\bf J} {\bf b} + {\bf b} {\bf J} {\bf c}$. The notation used in (\ref{Expansion}) for $\sqrt{\bf S}$ refers to the matrix
\begin{equation}
\sqrt{\bf S} := \left( \begin{array}{cc} {\bf J} & 0 \\ 0 & {\bf J} \end{array} \right) \left( \begin{array}{cc} {\bf a} & {\bf b} \\ {\bf b}^T & {\bf c} \end{array} \right).
\end{equation}

As can be seen from the expansion (\ref{Expansion}), in order to obtain the expression for ${\bf M}$ we need first to determine ${\bf S}^n$ and then, we have to insert the expression for ${\bf S}^n$ in (\ref{Expansion}) and calculate both sums therein. Let us  proceed in the next subsection with the first step: the calculation of ${\bf S}_{(n)}$. 

\subsection{Calculation of ${\bf S}_{(n)}$}

The matrix ${\bf S}$ is formed by four block $2 \times 2$ matrices where the upper left and the lower right are multiples of the identity matrix ${\bf 1}_{2\times2}$. The upper right block is the matrix ${\bf J} {\bf d}$ whereas the lower left is $- {\bf J} {\bf d}^T$. Notably, we found that this block structure is preserved after exponentiating the matrix ${\bf S}$ an integer number of times. That is to say, the n-power of matrix ${\bf S}$ gives a new matrix ${\bf S}^n$ given as
\begin{equation}
{\bf S}^n = \left( \begin{array}{cc} \alpha_n{\bf 1}_{2\times2} & \beta_n {\bf J} {\bf d} \\ - \beta_n {\bf J} {\bf d}^T & \gamma_n {\bf 1}_{2\times2} \end{array}\right). \label{MatrixSDef}
\end{equation} 
\noindent The coefficients $\alpha_n$, $\beta_n$ and $\gamma_n$, to be determined, depend on the values of the matrices ${\bf a}$, ${\bf b}$, ${\bf c}$ and ${\bf d}$. For $n=1$, these coefficients are given by the factors in the block matrices of ${\bf S} $ in  (\ref{SDefinition}) and can be directly defined as
\begin{equation}
\alpha_1 := - (\det {\bf a} + \det {\bf b}), \qquad \beta_1 := +1, \qquad \gamma_1 := - (\det {\bf c} + \det {\bf b}). \label{initialvalues}
\end{equation}

To calculate these coefficients for arbitrary $n$, first note that they can be generated with a linear operator ${\bf T}$ as
\begin{eqnarray}
\left( \begin{array}{c} \alpha_n \\ \beta_n \\ \gamma_n \end{array}\right) = {\bf T}^{n-1} \left( \begin{array}{c} \alpha_1 \\ \beta_1 \\ \gamma_1 \end{array}\right), \label{EqForCoeff}
\end{eqnarray}
\noindent where the matrix ${\bf T}$ is given by
\begin{equation}
{\bf T} = \left( \begin{array}{ccc} \alpha_1 & \beta_1 \det {\bf d} & 0 \\ \beta_1 & \gamma_1 & 0 \\ 0 & \beta_1 \det {\bf d} & \gamma_1 \end{array}\right).
\end{equation}

The calculation shows that the $n-1$ power of ${\bf T}$ is a matrix of the form
\begin{equation}
{\bf T}^{n-1} = \left( \begin{array}{cc} {\bf U}^{n-1} & \vec{0}^T \\ \vec{u}^T \gamma_1^{n-2} \sum^{n-2}_{j=0} \gamma^{-j}_1 {\bf U}^j & \gamma^{n-1}_1 \end{array}\right),
\end{equation}
\noindent where $\vec{0} = (0,0)$ and $\vec{u} = (0,  \beta_1 \det d)$ and matrix ${\bf U}$ is given by
\begin{eqnarray}
{\bf U} = \left( \begin{array}{cc} \alpha_1 & \beta_1 \det {\bf d} \\ \beta_1 & \gamma_1\end{array}\right).
\end{eqnarray}

Then, using (\ref{EqForCoeff}) we have the following relation for the coefficients
\begin{eqnarray}
\left( \begin{array}{c} \alpha_n \\ \beta_n \end{array}\right) = {\bf U}^{n-1} \left( \begin{array}{c} \alpha_1 \\ \beta_1 \end{array}\right), \qquad \gamma_n = \gamma^n_1 + \vec{u}^T \gamma^{n-2}_1 \sum^{n-2}_{j=0} \gamma^{-j}_1 {\bf U}^j \left( \begin{array}{c} \alpha_1 \\ \beta_1 \end{array}\right). \label{FExpressionCoeff}
\end{eqnarray}

In order to calculate ${\bf U}^{n-1}$ we need to diagonalize matrix ${\bf U}$ hence, let ${\bf P}$ be the matrix diagonalizing ${\bf U}$, then 
\begin{equation}
{\bf U} = {\bf P} \, {\bf D} \, {\bf P}^{-1}, \label{EqForU}
\end{equation}
\noindent where the matrix ${\bf P}$ is
\begin{equation}
{\bf P} = \left( \begin{array}{cc} \frac{(\lambda_+ - \gamma_1)}{\beta_1 } k_1 & \frac{(\lambda_{-} - \gamma_1)}{\beta_1} k_2 \\ k_1 & k_2 \end{array}\right).
\end{equation}

The real arbitrary parameters $k_1$ and $k_2$ result from the diagonalization procedure. Its values will be automatically cancelled as part of the calculation of ${\bf U}^{n-1}$ further below. The eigenvalues of ${\bf U}$, denoted by $\lambda_{\pm}$, have the following expression 
\begin{equation}
\lambda_{\pm} = \frac{\alpha_1 + \gamma_1}{2} \pm \frac{1}{2} \sqrt{ (\alpha_1 + \gamma_1)^2 - 4 (\alpha_1 \gamma_1 - \beta^2_1 \det {\bf d})}, \label{eigenvalues}
\end{equation}
\noindent and the diagonal matrix ${\bf D}$ is
\begin{equation}
{\bf D} = \left( \begin{array}{cc} \lambda_+ & 0 \\ 0 & \lambda_{-} \end{array}\right).
\end{equation}

We now take the $n-1$ power of ${\bf U}$ given in (\ref{EqForU}) to obtain the following result
\begin{eqnarray}
{\bf U}^{n-1} = \left( \begin{array}{cc} \frac{(\lambda_+ - \gamma_1)}{\beta_1 } k_1 & \frac{(\lambda_{-} - \gamma_1)}{\beta_1} k_2 \\ k_1 & k_2 \end{array}\right) \left( \begin{array}{cc} \lambda^{n-1}_+ & 0 \\ 0 & \lambda^{n-1}_{-} \end{array}\right) \left( \begin{array}{cc} \frac{(\lambda_+ - \gamma_1)}{\beta_1 } k_1 & \frac{(\lambda_{-} - \gamma_1)}{\beta_1} k_2 \\ k_1 & k_2 \end{array}\right)^{-1},
\end{eqnarray}
\noindent which, when combined with the result in (\ref{FExpressionCoeff}) together with the expression for $\vec{u}$, gives
\begin{eqnarray}
\alpha_n &=& \frac{1}{\sqrt{ (\alpha_1 - \gamma_1)^2 + 4 \beta^2_1 \det {\bf d} } } \left[ ( \lambda_+ - \gamma_1 ) \lambda^n_+ - ( \lambda_- - \gamma_1 ) \lambda^n_-   \right] ,\label{alpha} \\
\beta_n &=& \frac{1}{\sqrt{ (\alpha_1 - \gamma_1)^2 + 4 \beta^2_1 \det {\bf d} } } \left[ \lambda^n_+ -  \lambda^n_-   \right] ,\label{beta} \\
\gamma_n &=& \frac{1}{\sqrt{ (\alpha_1 - \gamma_1)^2 + 4 \beta^2_1 \det {\bf d} } } \left[ ( \lambda_+ - \gamma_1 ) \lambda^n_- - ( \lambda_- - \gamma_1 ) \lambda^n_+   \right]. \label{gamma}
\end{eqnarray}
\noindent These are the final expressions for the coefficients in ${\bf S}^n$. We are now ready to move to the second step: the analysis of the infinite series in (\ref{Expansion}).

\subsection{Series analysis}

Using the expression for the $n$ power of matrix ${\bf S}$, defined in (\ref{MatrixSDef}), the expression (\ref{Expansion}) can be written as
\begin{eqnarray}
{\bf M} &=& {\bf 1} + \sum^{+\infty}_{n=1} \frac{1}{(2n)!} \left( \begin{array}{cc} \alpha_n & \beta_n {\bf J} {\bf d} \\ - \beta_n {\bf J} {\bf d}^T & \gamma_n \end{array}\right) +  \sqrt{{\bf S}}\left[ {\bf 1} + \sum^{+\infty}_{n=1} \frac{1}{(2n+1)!}  \left( \begin{array}{cc} \alpha_n & \beta_n {\bf J} {\bf d} \\ - \beta_n {\bf J} {\bf d}^T & \gamma_n \end{array}\right) \right].
\end{eqnarray}
\noindent After collecting the components of each block matrix we obtain the following coefficients
\begin{eqnarray}
\alpha^{(e)} := 1 + \sum^{+\infty}_{n=1} \frac{1}{(2n)!} \alpha_n, \qquad \beta^{(e)} := \sum^{+\infty}_{n=1} \frac{1}{(2n)!} \beta_n, \qquad \gamma^{(e)} := 1 + \sum^{+\infty}_{n=1} \frac{1}{(2n)!} \gamma_n, \label{EvenCoeff}\\
\alpha^{(o)} := 1 + \sum^{+\infty}_{n=1} \frac{1}{(2n+1)!} \alpha_n, \qquad \beta^{(o)} := \sum^{+\infty}_{n=1} \frac{1}{(2n+1)!} \beta_n, \qquad \gamma^{(o)} := 1 + \sum^{+\infty}_{n=1} \frac{1}{(2n+1)!}\gamma_n. \label{OddCoeff}
\end{eqnarray}
\noindent Using this definition, the matrix ${\bf M}$ takes the following form 
\begin{equation}
{\bf M}({\bf a}, {\bf b}, {\bf c}) = \left( \begin{array}{cc}  \alpha^{(e)} {\bf 1}_{2\times2} &  \beta^{(e)} {\bf J} {\bf d} \\ - \beta^{(e)} {\bf J} {\bf d}^T & \gamma^{(e)} {\bf 1}_{2\times2} \end{array}\right) + \sqrt{{\bf S}}  \left( \begin{array}{cc}  \alpha^{(o)} {\bf 1}_{2 \times 2} & \beta^{(o)} {\bf J} {\bf d} \\ - \beta^{(o)} {\bf J} {\bf d}^T &  \gamma^{(o)} {\bf 1}_{2 \times 2} \end{array}\right) . \label{FinalformM}
\end{equation}

We now insert (\ref{alpha}), (\ref{beta}) and (\ref{gamma}) in the relations (\ref{EvenCoeff}) - (\ref{OddCoeff}) to obtain
\begin{eqnarray}
 \alpha^{(e)} &=&  \frac{1}{\sqrt{ (\det{\bf a} - \det{\bf c})^2 + 4 \det {\bf d} } } \left[ (\lambda_+ - \gamma_1) \cosh\sqrt{\lambda_+} - (\lambda_- - \gamma_1) \cosh\sqrt{\lambda_- } \right], \label{alphapar} \\
 \alpha^{(o)} &=& \frac{1}{\sqrt{ (\det{\bf a} - \det{\bf c})^2 + 4 \det {\bf d}  } } \left[ (\lambda_+ - \gamma_1) \frac{ \sinh\sqrt{\lambda_+}}{\sqrt{\lambda_+}}  - (\lambda_- - \gamma_1) \frac{ \sinh\sqrt{\lambda_- }}{\sqrt{\lambda_-}}  \right], \label{alphaimpar} \\
\beta^{(e)} &=& \frac{1}{\sqrt{ (\det{\bf a} - \det{\bf c})^2 + 4 \det {\bf d}  } } \left[  \cosh\sqrt{\lambda_+} -  \cosh\sqrt{\lambda_- } \right], \label{betapar} \\
\beta^{(o)} &=& \frac{1}{\sqrt{ (\det{\bf a} - \det{\bf c})^2 + 4 \det {\bf d}  } } \left[ \frac{ \sinh\sqrt{\lambda_+} }{\sqrt{\lambda_+}} - \frac{ \sinh\sqrt{\lambda_- }}{\sqrt{\lambda_-}} \right], \label{betaimpar} \\
 \gamma^{(e)} &=& \frac{1}{\sqrt{ (\det{\bf a} - \det{\bf c})^2 + 4 \det {\bf d}  } } \left[ (\lambda_+ - \gamma_1) \cosh\sqrt{\lambda_-} - (\lambda_- - \gamma_1) \cosh\sqrt{\lambda_+ } \right], \label{gammapar} \\
 \gamma^{(o)} &=& \frac{1}{\sqrt{(\det{\bf a} - \det{\bf c})^2 + 4 \det {\bf d}  } } \left[ (\lambda_+ - \gamma_1) \frac{ \sinh\sqrt{\lambda_-}}{\sqrt{\lambda_-}} - (\lambda_- - \gamma_1) \frac{\sinh\sqrt{\lambda_+ }}{\sqrt{\lambda_+}} \right], \label{gammaimpar}
\end{eqnarray}
\noindent where we have to recall the notation given in (\ref{initialvalues}) for $\alpha_1$, $\beta_1$ and $\gamma_1$ and the expression for the eigenvalues $\lambda_\pm$ in (\ref{eigenvalues}). These expressions link the components of the Lie algebra element ${\bf m}$ (RHS of \ref{FinalformM}) with the corresponding symplectic matrix ${\bf M}$ (LHS of ( \ref{FinalformM})) and constitute the main result of this section.

\subsection{Particular cases}
Let us consider in this subsection some particular cases which may be useful for some systems. First, notice that the general expression for the matrix components is of the form
\begin{eqnarray}
{\bf A} &=& \alpha^{(e)} + (\alpha^{(o)} - \beta^{(o)} \det{\bf b}) \, {\bf J} \, {\bf a} + \beta^{(o)} {\bf J} \,{\bf b} \, {\bf J} \, {\bf c}\, {\bf J} \, {\bf b}^T, \\
{\bf B} &=& (\gamma^{(o)} - \beta^{(o)} \det{\bf a}) \, {\bf J} \, {\bf b} + \beta^{(e)} ( {\bf J} \,{\bf a} \, {\bf J} \, {\bf b} +  {\bf J} \,{\bf b} \, {\bf J} \, {\bf c} )+  \beta^{(o)}  {\bf J} \,{\bf a} \, {\bf J} \, {\bf b} \, {\bf J} \, {\bf c}, \\
{\bf C} &=& (\alpha^{(o)} - \beta^{(o)} \det{\bf c}) \, {\bf J} \, {\bf b}^T + \beta^{(e)} ( {\bf J} \,{\bf b}^T \, {\bf J} \, {\bf a} +  {\bf J} \,{\bf c} \, {\bf J} \, {\bf b}^T )+  \beta^{(o)}  {\bf J} \,{\bf c} \, {\bf J} \, {\bf b}^T \, {\bf J} \, {\bf a}, \\
{\bf D} &=& \gamma^{(e)} + (\gamma^{(o)} - \beta^{(o)} \det{\bf b}) \, {\bf J} \, {\bf c} + \beta^{(o)} {\bf J} \,{\bf b}^T \, {\bf J} \, {\bf a}\, {\bf J} \, {\bf b}, 
\end{eqnarray}
\noindent which shows the non-linear relation between the Lie algebra elements, particularly the relation with Lie elements ${\bf b}$, and the symplectic group.

Let us consider the Lie element with ${\bf b}={\bf 0}$ and ${\bf a}, {\bf c} \neq 0$. In this case, the eigenvalues take the form $\lambda_+ = - \det{\bf a}$ and $\lambda_- = - \det{\bf c}$. After inserting ${\bf b}=0$ in (\ref{alphapar})-(\ref{gammaimpar}) we obtain the following symplectic matrix 
\begin{equation}
{\bf M}({\bf a}, {\bf b} = {\bf 0}, {\bf c}) = \left( \begin{array}{cc} \cosh\sqrt{- \det{\bf a}} + \frac{ \sinh\sqrt{- \det{\bf a}}}{\sqrt{-\det{\bf a}}}  \, {\bf J} \, {\bf a} & {\bf 0} \\ {\bf 0} & \cosh\sqrt{- \det{\bf c}} + \frac{ \sinh\sqrt{- \det{\bf c}}}{\sqrt{-\det{\bf c}}}  \, {\bf J} \, {\bf c} \end{array}\right). \label{Casebzero}
\end{equation}
\noindent Both block matrices in (\ref{Casebzero}) are elements of the group $Sp(2, \mathbb{R})$ which implies that the Lie algebra elements given by the parameters ${\bf a}$ and ${\bf c}$ can be considered as the Lie algebra generators of  $Sp(2, \mathbb{R})\otimes Sp(2, \mathbb{R}) \subset  Sp(4, \mathbb{R})$. 

Let us now consider the Lie algebra element with ${\bf a} = {\bf c} = 0$ and ${\bf b} \neq 0$. The eigenvalues are now given as $\lambda_+ = \lambda_- = - \det{\bf b}$ and the symplectic matrix takes the form
\begin{equation}
{\bf M}({\bf a} = {\bf 0}, {\bf b} , {\bf c}= {\bf 0}) = \left( \begin{array}{cc} \cosh\sqrt{- \det{\bf b}}  & \frac{\sinh\sqrt{- \det{\bf b}}}{\sqrt{- \det{\bf b}}}   \, {\bf J} \, {\bf b} \\  \frac{\sinh\sqrt{- \det{\bf b}}}{\sqrt{- \det{\bf b}}}   \, {\bf J} \, {\bf b}^T & \cosh\sqrt{- \det{\bf b}}  \end{array}\right). \label{Caseaandczero}
\end{equation}
\noindent This expression will be use in the next section to derive the classical squeeze operator. 

The matrices labelled by each of the ten parameters $a_{11}$, $a_{12}$, $a_{22}$, $b_{11}$, $b_{12}$, $b_{21}$, $b_{22}$, $c_{11}$, $c_{12}$ and $c_{22}$ can be considered as generators of the symplectic group $Sp(4, \mathbb{R})$. Notably, these are rather different to those generators reported for instance in \cite{milburn1984multimode}. The novel aspect of this result is that these generators are explicitly related with the Lie algebra generators. 

We are now ready to derive in the next section the relation between the squeeze operator and its corresponding symplectic matrix.

\section{Squeeze operators analysis} \label{ClassicalAnalysis}

The squeeze operator $\widehat{S}(\zeta)$ for a bi-partite system is of the form
\begin{equation}
\widehat{S}(\zeta) = e^{ \frac{1}{2} \left( \zeta^* \widehat{a}_1 \widehat{a}_2 - \zeta \widehat{a}^\dagger_1 \widehat{a}^\dagger_2 \right) }, \label{SqOperator}
\end{equation}
\noindent where $\widehat{a}_1$ and $\widehat{a}_2$ are the annihilation operators for the sub-systems, say, 1 and 2, of the bi-partite system. $\widehat{a}^\dagger_1$ and $\widehat{a}^\dagger_2$ are their adjoint operators respectively and $\zeta$ is a complex number labelling the amount of squeezing. This operator, when acting on the vacuum state of the bi-partite quantum harmonic oscillators, gives a family of squeezed states labelled by $\zeta$. 

As can be seen, the operator in  (\ref{SqOperator}) is given in the Fock representation. However, our analysis will be given in the Schr\"odinger representation which is described with operators $\widehat{q}_1$, $\widehat{q}_2$, $\widehat{p}_1$ and $\widehat{p}_2$. The relation between these representations is given by
\begin{eqnarray}
\widehat{a}_j =  \frac{1}{\sqrt{2}} \frac{\widehat{q}_j}{l_j} + \frac{i  }{\sqrt{2}}  \frac{ l_j \widehat{p}_j}{\hbar} , \qquad \widehat{a}^\dagger_j =  \frac{1}{\sqrt{2}} \frac{\widehat{q}_j}{l_j} - \frac{i  }{\sqrt{2}}  \frac{ l_j \widehat{p}_j}{\hbar} ,
\end{eqnarray}
\noindent for $j=1,2$ and $l_j := \sqrt{\frac{\hbar}{m_j \omega_j}}$ where $m_j$ and $\omega_j$ stand for the masses and the frequencies of the oscillators. Inserting these expressions for $\widehat{a}_j$ and $\widehat{a}^\dagger_j$ in the expression for $\widehat{S}(\zeta)$ the generator takes the following form
\begin{eqnarray}
\widehat{s}(\zeta) &:=& \frac{1}{2} \left( \zeta^* \widehat{a}_1 \widehat{a}_2 - \zeta \widehat{a}^\dagger_1 \widehat{a}^\dagger_2 \right) , \nonumber \\
&=& \frac{1}{2i \hbar} \left[  \frac{\hbar \, \zeta_y}{l_1 l_2} \; \widehat{q}_1 \widehat{q}_2 -  \frac{l_2 \, \zeta_x}{l_1} \; \widehat{q}_1 \widehat{p}_2 -  \frac{l_1\, \zeta_x}{l_2} \; \widehat{p}_1 \widehat{q}_2 -  \frac{l_1 l_2 \, \zeta_y}{\hbar} \widehat{p}_1 \widehat{p}_2  \right], \label{SqGen}
\end{eqnarray}
\noindent where $\zeta_x$ and $\zeta_y$ are the real and imaginary parts of $\zeta$. 

The expression in (\ref{SqGen}) can be written in the form
\begin{equation}
\widehat{s}(\zeta) = - \frac{i}{4\hbar} (\vec{\widehat{R}}^T_1, \vec{\widehat{R}}^T_2) \left( \begin{array}{cc} 0 & {\bf b} \\ {\bf b}^T & 0 \end{array}\right) \left( \begin{array}{c} \vec{\widehat{R}}_1 \\ \vec{\widehat{R}}_2 \end{array}\right), \label{SqueezedGen}
\end{equation}
\noindent where the matrix ${\bf b}$ takes the form
\begin{equation}
{\bf b} = \left( \begin{array}{cc}  \frac{\hbar \, \zeta_y}{l_1 l_2} & - \frac{l_2 \,  \zeta_x}{l_1}  \\ -  \frac{l_1\, \zeta_x}{l_2}  & -  \frac{l_1 l_2 \, \zeta_y}{\hbar}  \end{array}\right). \label{bMatrix}
\end{equation}

The Lie algebra $sp(4, \mathbb{R})$ and the Lie algebra of second order polynomials with Poisson bracket given by the commutator are isomorphic \cite{hall2018theory, adesso2014continuous}. This relation is given by the following isomorphism
\begin{equation}
{\bf m} = \left( \begin{array}{cc} {\bf J} & 0 \\ 0 & {\bf J} \end{array} \right) {\bf L} \qquad \Leftrightarrow \qquad - \frac{i}{4\hbar} (\vec{\widehat{R}}^T_1, \vec{\widehat{R}}^T_2) {\bf L} \left( \begin{array}{c} \vec{\widehat{R}}_1 \\ \vec{\widehat{R}}_2 \end{array}\right), \label{Isomorphism}
\end{equation}
\noindent and can be checked that the Lie algebra multiplication is preserved under (\ref{Isomorphism}). Using this ismorphism, the squeezed generator in (\ref{SqueezedGen}) maps to an element ${\bf m}_{\bf s} \in sp(4,\mathbb{R})$ given by
\begin{equation}
{\bf m}_{\bf s} = \left( \begin{array}{cc} {\bf J} & 0 \\ 0 & {\bf J} \end{array} \right)  {\bf L}_{\bf s} ,
\end{equation}
\noindent where ${\bf L}_{\bf s} $ is given as
\begin{equation}
{\bf L}_{\bf s}  = \left( \begin{array}{cc} 0 & {\bf b} \\ {\bf b}^T & 0 \end{array}\right),
\end{equation}
\noindent and matrix ${\bf b}$ is given in (\ref{bMatrix}). The index ${\bf s}$ stands for the relation of these elements with the squeeze operator $\widehat{s}(\zeta)$ in (\ref{SqGen}) .

We are now ready to calculate the symplectic matrix ${\bf M}_{\bf s} $ given by the exponential 
\begin{equation}
{\bf M}_{\bf s}(r, \phi) = e^{  \left( \begin{array}{cc} {\bf J} & 0 \\ 0 & {\bf J} \end{array} \right) \left( \begin{array}{cc} 0 & {\bf b} \\ {\bf b}^T & 0 \end{array}\right)  },
\end{equation}
\noindent and recall that $\zeta_x$ and $\zeta_y$, are related with the components of the matrix ${\bf b}$ via (\ref{bMatrix}). 

The expression for matrix ${\bf M}_{\bf s}(r, \phi)$ results from inserting (\ref{bMatrix}) in (\ref{Caseaandczero}). After this substitution, it yields the following matrix
\begin{equation}
{\bf M}_{\bf s}(r, \phi) = \left(
\begin{array}{cccc}
 \cosh (r) & 0 & -\sinh (r) \cos (2 \phi ) \frac{l_1}{l_2}& - \sinh (r) \sin (2 \phi ) \frac{l_1 l_2}{\hbar}\\
 0 & \cosh (r) & -\sinh (r) \sin (2 \phi ) \frac{\hbar}{l_1 l_2} & \sinh (r) \cos (2 \phi ) \frac{l_2}{l_1} \\
 -\sinh (r) \cos (2 \phi ) \frac{l_2}{l_1} & - \sinh (r) \sin (2 \phi ) \frac{l_1 l_2}{\hbar} & \cosh(r) & 0 \\
-\sinh (r) \sin (2 \phi ) \frac{\hbar}{l_1 l_2} & \sinh (r) \cos (2 \phi ) \frac{l_1}{l_2} & 0 & \cosh (r) \\
\end{array} 
\right), \label{FinalFormMS}
\end{equation}
\noindent  where $r$ and $\phi$ are related with the squeezing parameter as $\zeta = r e^{2 i \phi}$. 

The main result of this subsection is the expression for ${\bf M}_{\bf s}$, which we call, classical squeeze operator. Its squeezing properties will be explore in the next subsection.

Note that despite the Planck's constant appears in the components of ${\bf M}_{\bf s}(r, \phi) $ it is absorbed by the factors $l_1 l_2$. As a result, any connection with the quantum squeeze operator is removed, i.e., ${\bf M}_{\bf s}$ is strictly a classical operator on the classical phase space of the system. Moreover, it is worth to mention that more general symplectic group elements can be derived using the expressions (\ref{alphapar})-(\ref{gammaimpar}). In such cases, we only need the corresponding Lie algebra matrix ${\bf m}$ that rises from the isomorphism (\ref{Isomorphism}). However, in the present case, we only focused on those related with the squeeze operator $\widehat{s}(\zeta) $.

The next subsection will show the main features of matrix ${\bf M}_{\bf s}$ as part of the $Sp(4,\mathbb{R})$ group.

\subsection{Classical analysis}

The aim of this subsection is to check whether the canonical transformation ${\bf M}_{\bf s}$ can be considered as a squeezing classical operator. Due to the hyperbolic functions in the components of ${\bf M}_{\bf s}$ this seems to be the case.

In the quantum scenario, the correlation matrix ${\bf V}^{(2)}$ \cite{braunstein2005quantum} arising from the squeezed state associated with the operator in (\ref{SqOperator}) is of the form
\begin{equation}
{\bf V}^{(2)} = \frac{1}{4}\left(
\begin{array}{cccc}
 \cosh (2 r) & 0 & \sinh (2 r)  & 0 \\
 0 & \cosh (2 r) & 0 & - \sinh (2 r)  \\
 \sinh (2 r) & 0 & \cosh (2 r) & 0 \\
 0 & - \sinh (2 r)  & 0 & \cosh (r) \\
\end{array}
\right). \label{V2Matrix}
\end{equation}
\noindent  Its components ${ V}^{(2)}_{i j}$ are given as
\begin{equation}
{ V}^{(2)}_{i j} =  \frac{1}{2} \langle \Psi_s | \widehat{\xi}_i \widehat{\xi}_j + \widehat{\xi}_j \widehat{\xi}_i | \Psi_s \rangle,
\end{equation}
\noindent where $| \Psi_s \rangle = \widehat{S}(\zeta) | 0\rangle_1 | 0 \rangle_2$ is the squeezed state and $\widehat{\xi} = (\widehat{q}_1, \widehat{p}_1, \widehat{q}_2, \widehat{p}_2)$ and $\widehat{q}_j$, $\widehat{p}_j$, $j=1,2$ are the quadratures of the system. The state $ | 0\rangle_1 | 0 \rangle_2$ is the vacuum state of the two quantum harmonic oscillators.

We can note the similitude between these two matrices ${\bf V}^{(2)} $ and ${\bf M}_{\bf s}$. To make the similitude more explicit, let us consider $\phi=\pi/2$ and $l_1 = l_2$ and $m_j  \omega_j =1$ for $j=1,2$. With these conditions the matrix ${\bf M}_{\bf s}$ takes the following form
\begin{equation}
{\bf M}_{\bf s}\left(r, \frac{\pi}{2}\right) = \left(
\begin{array}{cccc}
 \cosh (r) & 0 & \sinh (r)  & 0 \\
 0 & \cosh (r) & 0 & -\sinh (r) \\
 \sinh (r)  & 0 & \cosh(r) & 0 \\
0 & -\sinh (r)  & 0 & \cosh (r) \\
\end{array}
\right).
\end{equation}

Remarkably, only two differences arise: (i) the global factor in (\ref{V2Matrix}) and (ii) the factor 2 in the argument of the hyperbolic functions. This factor implies that the amount of squeezing in ${\bf V}^{(2)} $ doubles the amount of squeezing in ${\bf M}_{\bf s}$.  That is to say, for each value of the squeezing parameter $r$, the amount of squeezing of ${\bf V}^{(2)} $ is twice the amount of squeezing of ${\bf M}_{\bf s}\left(r, \frac{\pi}{2}\right)$.

\begin{figure}[bt!]
     \centering
     \begin{subfigure}[b]{0.45\textwidth}
         \centering
         \includegraphics[width=0.4\textwidth]{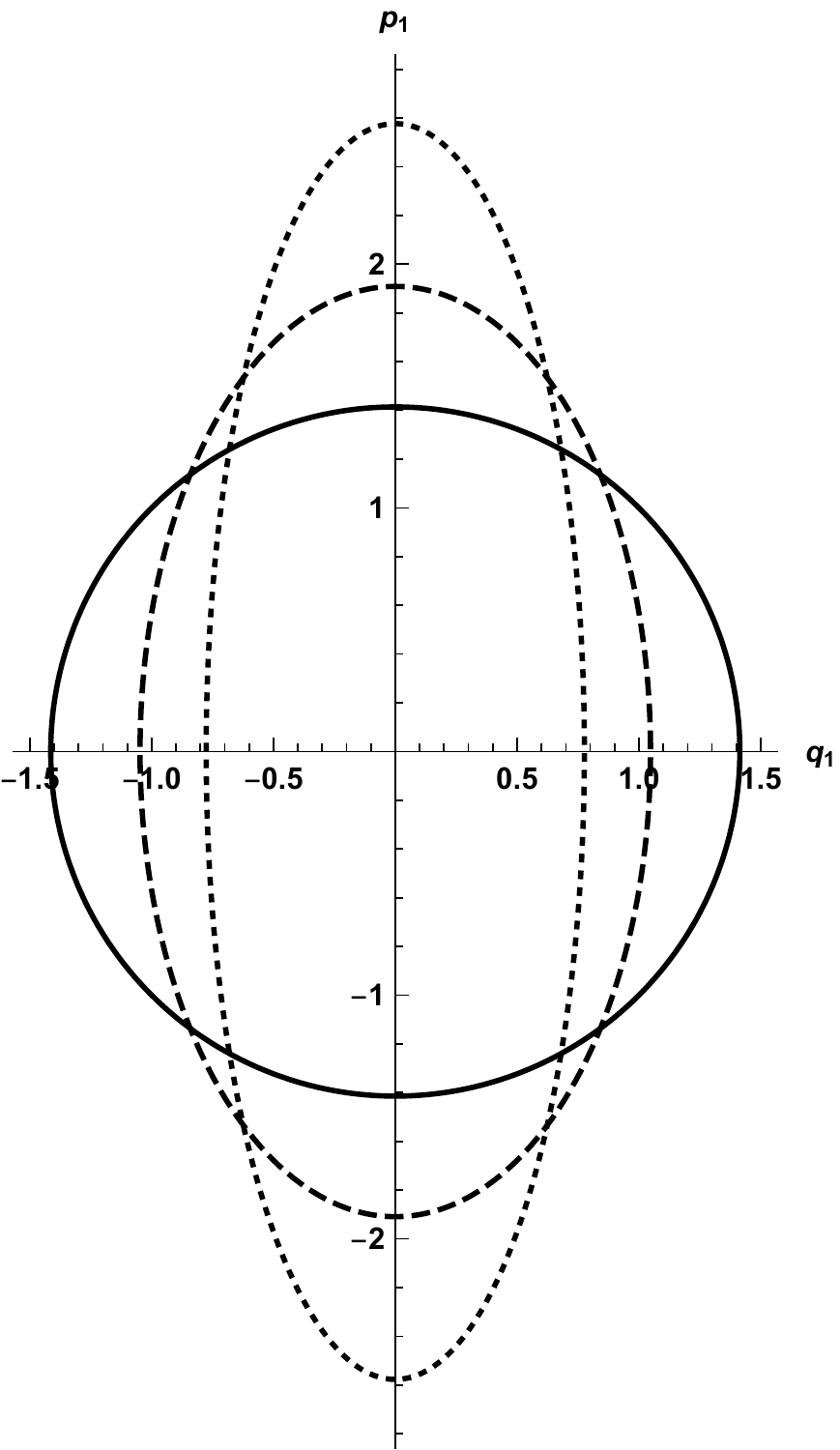}
         \caption{{\footnotesize{Squeezing a circular trajectory.}}}
         \label{Squeezing}
     \end{subfigure}
     \hspace{0.5cm}
     \begin{subfigure}[b]{0.45\textwidth}
         \centering
         \includegraphics[width=0.65\textwidth]{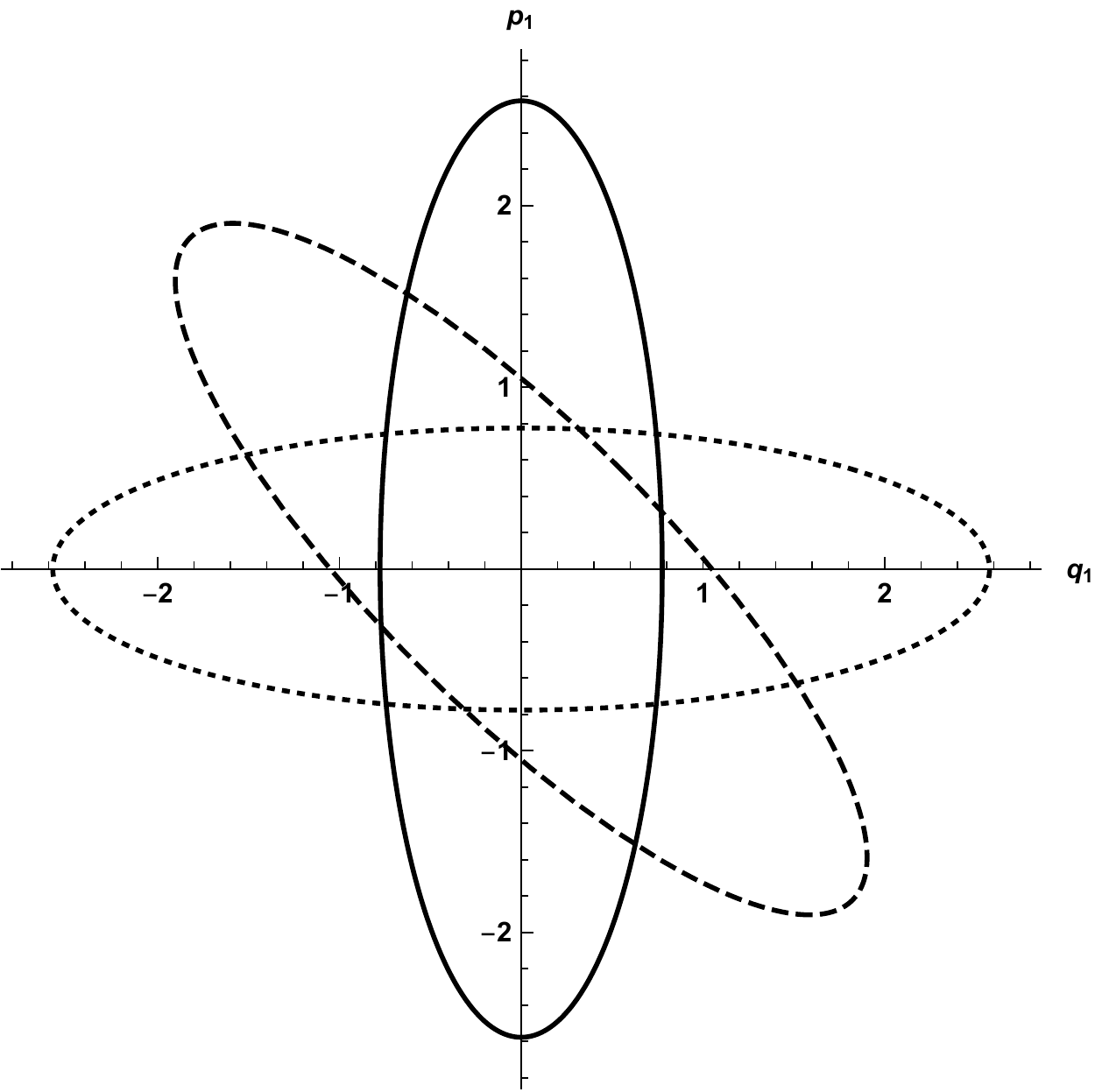}
         \caption{{\footnotesize{Rotating a squeezed trajectory with $r=0.6$. }}}
         \label{Rotation}
     \end{subfigure}
        \caption{{\footnotesize{ In both figures, the solid, the dashed and the dotted lines correspond to: in (a) $r=0$, $r=0.3$ and $r=0.6$ respectively and in (b) to $\phi=0$, $\phi=\pi/4$ and $\phi=\pi/2$ respectively.}}}
        \label{Figure}
\end{figure}

Finally, in figure \ref{Figure} we show the squeezing and rotation properties of the matrix ${\bf M}_{\bf s}$ as a canonical transformation for different values of $r$ and $\phi$. In this figure we consider the action of ${\bf M}_{\bf s}$ on a circular trajectory $(q_1(t), p_1(t), q_2(t), p_2(t))$ were, $q_j(t) = \cos(t) \; q_j + \sin(t) \; p_j$ and $p_j(t) = - \sin(t) \; q_j + \cos(t) \; p_j$, for $j=1,2$. The action of ${\bf M}_{\bf s}$ is explicitly of the form
\begin{equation}
\left( \begin{array}{c} q'_1(t) \\ p'_1(t) \\ q'_2(t) \\ p'_2(t) \end{array} \right) = {\bf M}_{\bf s}\left(r, \phi \right) \left( \begin{array}{c} q_1(t) \\ p_1(t) \\ q_2(t) \\ p_2(t) \end{array} \right).
\end{equation}

In \ref{Figure} we showed the graph of $(q'_1(t), p'_1(t))$. As expected, we note in \ref{Squeezing} that the amount of squeezing is labelled by $r$ and that $\phi$ labels the rotation as showed in \ref{Rotation}.

\section{Conclusions} \label{Conclusion}

In this paper we provided the direct relation between the elements of the Lie algebra $sp(4, \mathbb{R})$ and its group elements $Sp(4, \mathbb{R})$. This was given in (\ref{FinalformM}) together with the Eqs. (\ref{alphapar})-(\ref{gammaimpar}). As we mentioned in the introduction, this relation is particularly useful in calculations where the unitary representation of the symplectic group plays an important role rather than the Wigner representation scheme \cite{Arvind:1995ab} or the Fock representation \cite{ma1990multimode} analysis. Moreover, in (\ref{Casebzero}) and (\ref{Caseaandczero}) we derived the group generators in terms of the Lie algebra parameters. As far 

We also obtained the expression for the symplectic matrix (\ref{FinalFormMS}) whose unitary representation gives rise to the squeeze operator (\ref{SqOperator}). We showed that this symplectic matrix is also a squeezing transformation when considered at classical level. Remarkably, the classical squeeze matrix shares similarities with the correlation matrix (\ref{V2Matrix}) which suggests a deeper connection between both matrices which so far has not yet being stablished. This idea is reinforced if we notice that the amount of squeezing of ${\bf M}_{\bf s}$ is half the amount of squeezing of ${\bf V}^{(2)}$ in (\ref{V2Matrix}). 

\section{Acknowledgments}
I thank M. Berm\'udez-Monta\~na for her useful comments along the preparation of this work.


\begin{thebibliography}{99}
\bibitem{walls1983squeezed} Walls, D. F., ``Squeezed states of light'', \emph{Nature}, {\bf 306}, 5939, 1983, Springer.

\bibitem{schnabel2017squeezed} Schnabel, R., ``Squeezed states of light and their applications in laser interferometers'', \emph{Physics Reports}, {\bf 684}, 2017, Elsevier.

\bibitem{walls2007quantum} Walls, D. F. and Milburn, G. J., ``Quantum optics'', 2007, \emph{Springer Science \& Business Media}.

\bibitem{adesso2014continuous}   Adesso, G., Ragy, S. and Lee, A. R., ``Continuous variable quantum information: Gaussian states and beyond'', \emph{Open Systems \& Information Dynamics}, {\bf 21}, 2014, World Scientific.

\bibitem{braunstein2005quantum} Braunstein, S. L. and Van Loock, P., ``Quantum information with continuous variables'', \emph{Reviews of Modern Physics}, {\bf 77}, 2, 2005.

\bibitem{gasperini1993quantum} Gasperini, M. and Giovannini, M., ``Quantum squeezing and cosmological entropy production'', \emph{Classical and Quantum Gravity}, {\bf 10}, 9, 1993.

\bibitem{grain2019squeezing} Grain, J. and Vennin, V., ``Squeezing formalism and canonical transformations in cosmology'', \emph{arXiv preprint arXiv:1910.01916}, 2019.

\bibitem{ashtekar2003mathematical} Ashtekar, A., Bojowald, M. and Lewandowski, J., ``Mathematical structure of loop quantum cosmology'', \emph{Advances in Theoretical and Mathematical Physics}, {\bf 7}, 2, 2003.

\bibitem{bojowald2010canonical} Bojowald, M., ``Canonical gravity and applications: cosmology, black holes, and quantum gravity'', \emph{Cambridge University Press}, 2010.

\bibitem{bojowald2011quantum} Bojowald, M., ``Quantum cosmology: a fundamental description of the universe'', vol. {\bf 835}, \emph{Springer Science \& Business Media}, 2011.

\bibitem{mielczarek2012gaussian} Mielczarek, J. and Piechocki, W., ``Gaussian state for the bouncing quantum cosmology'', \emph{Physical Review D}, {\bf 86}, 8, 2012.

\bibitem{gazeau2013quantum} Gazeau, J. P., Mielczarek, J. and Piechocki, W., ``Quantum states of the bouncing universe'', \emph{Physical Review D}, {\bf 87}, 12, 2013.

\bibitem{diener2014numerical} Diener, P., Gupt, B., Megevand, M. and Singh, P., ``Numerical evolution of squeezed and non-Gaussian states in loop quantum cosmology'', \emph{Classical and Quantum Gravity}, {\bf 31}, 16, 2014.

\bibitem{ashtekar2006quantum} Ashtekar, A., Pawlowski, T. and Singh, P., ``Quantum nature of the big bang: an analytical and numerical investigation'', \emph{Physical Review D}, {\bf 73}, 12, 2006.

\bibitem{ashtekar2006quantum2} Ashtekar, A., Pawlowski, T. and Singh, P., ``Quantum nature of the big bang: Improved dynamics'', \emph{Physical Review D}, {\bf 74}, 8, 2006.

\bibitem{husain2019matter} Husain, V. and Singh, S., ``Matter-Geometry entanglement in quantum gravity'', arXiv: 1907.03776, 2019.

\bibitem{ashtekar2003quantum}  Ashtekar, A., Fairhurst, S. and Willis, J. L., ``Quantum gravity, shadow states and quantum mechanics'',  \emph{Classical and Quantum Gravity}, {\bf 20}, 6, 2003.

\bibitem{corichi2007polymer} Corichi, A., Vukasinac, T. and Zapata, Jos\'e A., ``Polymer quantum mechanics and its continuum limit'', Phys. Review D, {\bf 76}, 4, 2007.

\bibitem{flores2013propagators} Flores-Gonz\'alez, E. and Morales-T\'ecotl, H. A. and Reyes, J. D., ``Propagators in polymer quantum mechanics'', \emph{Annals of Physics}, {\bf 336}, 2013.

\bibitem{velhinho2007quantum} Velhinho, J. M., ``The quantum configuration space of loop quantum cosmology'', \emph{Classical and Quantum Gravity}, {\bf 24}, 14, 2007.

\bibitem{pawlowski2014separable}  Pawlowski, T., Villase\~nor, E. J. S.  and Barbero, J. F., ``Separable Hilbert space for loop quantization'', Phys. Review D, {\bf 90}, 6, 2014.

\bibitem{Garcia-Chung:2020cag}  A.~Garcia-Chung, ``The symplectic group in Polymer Quantum Mechanics,'' arXiv:2003.00388 [gr-qc].

\bibitem{hall2018theory} Hall, M., ``The theory of groups'', 2018, \emph{Courier Dover Publications}

\bibitem{moshinsky1971linear}  Moshinsky, M. and Quesne, C., ``Linear canonical transformations and their unitary representations'', \emph{Journal of Mathematical Physics}, {\bf 12}, 8, 1971.

\bibitem{wolf2016development} Wolf, K., ``Development of linear canonical transforms: a historical sketch'', in \emph{Linear Canonical Transforms}, 3-28, 2016, Springer.

\bibitem{torre2005linear}  Torre, A., ``Linear ray and wave optics in phase space: bridging ray and wave optics via the Wigner phase-space picture'', 2005, Elsevier.


\bibitem{Arvind:1995ab}  Arvind, B.~Dutta, N.~Mukunda and R.~Simon, ``The Real symplectic groups in quantum mechanics and optics,'' Pramana {\bf 45}, 471 (1995).

\bibitem{ma1990multimode} Ma, X. and Rhodes, W., ``Multimode squeeze operators and squeezed states'', \emph{Physical Review A}, {\bf 41}, 9, 1990.

\bibitem{milburn1984multimode} Milburn, G. J., ``Multimode minimum uncertainty squeezed states'', \emph{Journal of Physics A: Mathematical and General}, {\bf 17}, 4, 1984.

\end{thebibliography}
\end{document}